\documentclass[aps,amsfonts,amsmath,nofootinbib,byrevtex,twocolumn,showpacs,floatfix,final,prd,a4paper]{revtex4}
\usepackage{graphicx}

\newcommand{\be}{\begin{equation}}
\newcommand{\ee}{\end{equation}}
\newcommand{\braket}[2]{ \langle #1 | #2 \rangle }
\newcommand{\bra}[1]{\langle #1 |\,}
\newcommand{\ket}[1]{\,| #1 \rangle}
\DeclareMathOperator{\Det}{Det}				
\newcommand{\comm}[2]{ \big[ #1 , #2 \big]  }		
\renewcommand{\vec}[1]{\boldsymbol{\mathrm{#1}}} 	

\newcommand{\vx}{\vec{x}}
\newcommand{\vy}{\vec{y}}
\newcommand{\vp}{\vec{p}}
\newcommand{\vq}{\vec{q}}
\newcommand{\vk}{\vec{k}}
\newcommand{\uvx}{\Hat{\boldsymbol{\mathrm{x}}}}
\newcommand{\uvq}{\Hat{\boldsymbol{\mathrm{q}}}}
\newcommand{\uvk}{\Hat{\boldsymbol{\mathrm{k}}}}
\newcommand{\uvp}{\Hat{\boldsymbol{\mathrm{p}}}}

\renewcommand{\d}[1][]{\mathrm{d}^{#1}} 	
\newcommand{\dfr}[2][1]{\frac{\mathrm{d}^{#1}#2}{(2\pi)^{#1}}}
\newcommand{\calD}{\ensuremath{\mathcal{D}}}
\newcommand{\calJ}{\ensuremath{\mathcal{J}}}
\newcommand{\calL}{\ensuremath{\mathcal{L}}}
\newcommand{\calM}{\ensuremath{\mathcal{M}}}
\newcommand{\calN}{\ensuremath{\mathcal{N}}}

\begin{document}

\title{Topological susceptibility in $SU(2)$ Yang--Mills theory in the Hamiltonian approach in Coulomb gauge} 
\author{Davide R. Campagnari}
\author{Hugo Reinhardt}
\affiliation{Institut f\"ur Theoretische Physik, Universit\"at T\"ubingen, Auf der Morgenstelle 14, D-72076 T\"ubingen, Germany}

\begin{abstract}
The topological susceptibility is calculated within the Hamiltonian approach to Yang--Mills theory in Coulomb gauge, using the vacuum wave functional previously determined by a variational solution of the Yang--Mills Schr\"odinger equation. The numerical result agrees qualitatively with the predictions of lattice simulations.
\end{abstract} 


\maketitle


\section{Introduction}

At the classical level, quantum chromodynamics (QCD) with massless fermions is  chirally
symmetric; that is to say, the QCD Lagrangian is invariant under separate global flavor
rotations of the left- and right-handed quarks. For $N_f$ massless quark flavors
the chiral symmetry group is
\be
\label{G1}
SU_V(N_f) \times SU_A(N_f) \times U_V(1) \times U_A (1) \; ,
\ee
where the vector (axial vector) symmetry groups denoted by a subscript $V$ $(A)$
rotate left- and right-handed fermions in the same (opposite) way.
The chiral symmetry is a good starting point for the $N_f = 3$ light quark
flavors $u, d, s$. In the quantum theory, the $SU_A(N_f)$ is spontaneously
broken by the dynamical condensation of quarks, $\langle \bar{q} q \rangle \neq
0$, which results in the generation of a constituent quark mass and gives rise to
$N^2_f - 1$ Goldstone bosons, which can be identified with the octet of light
pseudoscalar mesons. Further, the small but finite (current) quark masses break
the axial $SU_A(N_f)$ explicitly and induce a mass for the pseudoscalar
mesons. Furthermore, the $SU_V(N_f)$ is softly broken by the differences in the
current quark masses, which lifts the mass degeneracy of the pseudoscalar
mesons. The $U_V(1)$ symmetry corresponds to baryon number conservation and
remains unbroken in the QCD vacuum. The $U_A(1)$ symmetry, however, has been
an issue for quite some time. If the $U_A(1)$ were intact, the light
hadrons would occur in degenerate parity doublets, which is not the case.
Furthermore, if $U_A(1)$ were spontaneously broken, there should be a (nearly)
massless (or at least light) pseudoscalar flavor singlet meson, the 
``would-be'' Goldstone boson of spontaneous $U_A(1)$ symmetry breaking. 
The only candidate is the $\eta'$, which, however, is by far too heavy to qualify for this
particle.

It was first shown by Adler \cite{Adl69}, Bell and Jackiw \cite{BelJac69} that the $U_A(1)$ is anomalously
broken, \textit{i.e.}, broken by quantum effects. The anomalous breaking of the global
$U_A(1)$ symmetry manifests itself in the non-invariance of the fermionic
integration measure \cite{Fuj79,Fuj80} and results in the well-known axial anomaly, which for massless quarks reads
\be
\label{G2}
\partial_\mu j^\mu_5 (x) = 2 \, N_f \, q (x) \; .
\ee
Here, $j^\mu_5 (x) = \bar{\psi} (x) \gamma^\mu \gamma_5 \psi (x)$ is the axial
current and
\be
\label{G3}
q(x) = \frac{g^2}{32 \pi^2} \: F^a_{\mu \nu} (x) \widetilde{F}^{\mu \nu}_a (x)
\ee
is the topological charge density in Minkowski space, with
\begin{subequations}
\label{G35}
\begin{align}
F_{\mu\nu}^a &= \partial_\mu A_\nu^a - \partial_\nu A_\mu^a + g \, f^{abc} A_\mu^b A_\nu^c \\
\widetilde{F}^{\mu \nu}_a &= \frac{1}{2} \; \varepsilon^{\mu \nu \rho \sigma} F^a_{\rho \sigma} 
\end{align}
\end{subequations}
being the field strength tensor and its dual ($f^{abc}$ are the structure constants of the $\mathfrak{su}(N_c)$ algebra and $g$ is the coupling constant; we use color sub- and superscripts indiscriminately). Adler and Bardeen showed \cite{AdlBar69} that Eq.~\eqref{G2} does not recieve corrections from higher-order diagrams. Although $q (x)$ is a total divergence
\be
F^a_{\mu \nu} \, \widetilde{F}^{\mu \nu}_a  = 2 \partial_\mu K^\mu ,
\ee
where
\be
K^\mu = \varepsilon^{\mu\nu\rho\sigma}
\left[ A_\nu^a \partial_\rho A_\sigma^a + \frac13 \, g f^{abc} \, A_\nu^a A_\rho^b A_\sigma^c \right]
\ee
is the topological current, there are topologically nontrivial field configurations
(in Euclidean space) such as instantons \cite{Pol75,Bel+75}, magnetic monopoles \cite{Rei97}
or center vortices \cite{Rei02a,EngRei00}, for which the topological charge
\be
\label{G4}
Q = \int \d[4]x_E \: q(x)
\ee
is non-zero (for smooth field configurations of finite Euclidean action, $Q$ is integer-valued).
As a consequence of the existence of these field configurations, the axial charge
\be
\label{G5}
Q_5 = \int \d[3]x \: j^0_5(x)
\ee
is not conserved \cite{tHo76,tHo76a}.

Using large-$N_c$ arguments, Witten \cite{Wit79} and Veneziano \cite{Ven79}
showed that the axial anomaly provides a mass term for the pseudoscalar flavor
singlet meson given by
\be
\label{G6}
m^2_{\eta'} + m^2_\eta - 2 m^2_K = \frac{2 N_f}{F^2_\pi} \: \chi \; ,
\ee
where $F_\pi \simeq 93$ MeV is the pion decay constant and 
\be
\label{G7}
\chi = \int \d[4]x_E \; \bra{0} q (x) \, q (0) \ket{0}
\ee
is the topological susceptibility, which is a purely gluonic quantity (in Minkowski space
Eq.~\eqref{G7} has an additional factor $-i$). Since this quantity is defined as vacuum
expectation value it is clear that its evaluation requires nonperturbative
methods. Indeed, in perturbation theory the topological susceptibility vanishes
to all orders. This quantity has been calculated on the lattice (see Refs.~\cite{DebGiuPic05,Dur+07} for recent calculations)
and the results are compatible with the prediction of the Witten--Veneziano formula
Eq.~(\ref{G6})
\be
\label{G8}
\chi \simeq ( 180 \mbox{ MeV} )^4
\ee
using the experimental data for the meson masses and $F_\pi$ as input.

Like the string tension \cite{Deb+97}, the topological susceptibility seems to be
dominated by center vortices \cite{BerEngFab01}. In fact, a center vortex model of the
infrared sector of Yang--Mills theory \cite{EngRei00a} yields a value for $\chi$
compatible with lattice results \cite{Eng00}.

Topologically nontrivial Euclidean field configurations such as instantons and
center vortices describe quantum tunneling between topologically different
Yang--Mills vacua \cite{Jac80}, see Eq.~\eqref{G9} below. Like in quantum mechanics, this quantum tunneling is fully accounted
for by the solution of the Yang--Mills Schr\"odinger equation.
Recently, progress has been made in determining the vacuum wave functional by a
variational solution of the Yang--Mills Schr\"odinger equation in Coulomb gauge
\cite{SzcSwa01,Szc04,FeuRei04gb,FeuRei04,ReiFeu05,EppReiSch07,ReiEpp07,FeuRei08,Epp+07}.
There is good evidence to believe that the obtained wave functional contains the essential infrared physics:
the absence of gluons from the physical spectrum in the infrared \cite{FeuRei04}, a linearly rising
potential for static color charges \cite{EppReiSch07} and a perimeter law for the
't Hooft loop \cite{ReiEpp07}. In the present paper we use the Yang--Mills
wave functional determined in refs.\ \cite{FeuRei04,EppReiSch07}  to calculate the topological
susceptibility given by Eq.~\eqref{G7}.

The organization of the paper is as follows: 
In the next section we briefly review the $\theta$-vacuum in the Hamiltonian
approach. In section 3 we derive the expressions for the topological
susceptibility in the Hamiltonian approach in Coulomb gauge and evaluate the relevant matrix
elements. Our numerical results are presented in section 4. Some concluding remarks are given in section 5.


\section{The $\theta$-vacuum in the canonical quantization approach}

Consider Yang--Mills theory in the Weyl gauge $A_0^a = 0$. The classical
(time-independent) vacuum configurations are pure gauge spatial fields
\be
\label{G9}
A_i = A_i^a \; t_a = \frac{i}{g} \; U \partial_i U^\dagger ,
\ee
where $t_a$ are the hermitean generators of the gauge group.
Imposing the usual boundary condition that the gauge function $U(\vx) \in SU(N_c)$
approaches a unique value for $| \vx | \to \infty$ (independent of the direction
$\uvx$) compactifies $\mathbb{R}^3$ to $S_3$ and consequently the $U(\vx)$ can be
classified according to the winding number $n [U] \in \Pi_3 (S_3)$ \cite{Jac80}. The
classical vacuum configurations Eq.~\eqref{G9} belonging to different winding
numbers are separated by infinite potential barriers. In the quantum theory,
tunnelling between the different classical vacua occurs and in a semiclassical
picture, the barrier penetration is described by instantons,
(space-)time-dependent solutions of the classical Euclidean Yang--Mills equation of motion,
which interpolate between vacuum configurations differing in the winding number by
$\pm 1$. We will not resort here to a semiclassical description but instead
approximately solve the Schr\"odinger equation, which fully accounts for the quantum tunnelling.

In Weyl gauge the physical coordinates are the spatial gauge fields
$A_i^a(x)$ and the corresponding canonical conjugate momenta are given by
the chromoelectric field $E_i^a(x) = F_{0i}^a(x)$. In canonical quantization the electric field is promoted
to the momentum operator \mbox{$\Pi_i^a(x)=-i \delta/\delta A_i^a(x)$} satisfying
canonical commutation relations, and the Yang--Mills Hamiltonian reads
\be
\label{G10}
H = \frac{1}{2} \int \d[3]x \: \left[ \big( \Pi^a_i(\vx) \big)^2 + \big( B^a_i(\vx) \big)^2 \right] ,
\ee
where
\be
\label{G11}
B^a_i = \frac12 \, \varepsilon_{ijk} \, F^a_{jk} = \varepsilon_{ijk} \left[ \partial_j A^a_k + \frac{g}{2} f^{a b c} A^b_j
A^c_k \right]
\ee
is the non-Abelian magnetic field. 

In Weyl gauge, Gauss' law is lost from the equations of motion and has to be
imposed as a constraint on the wave functional
\be
\label{G12}
\hat{D}^{ab}_i \, \Pi^b_i (\vx) \, \Psi [A] = - g \, \rho_\mathrm{ext}^a (\vx) \, \Psi [A] \; .
\ee
Here,
\be
\label{G13}
\hat{D}^{a b}_i = \delta^{a b} \partial_i + g \hat{A}^{a b}_i \; , \;
\hat{A}^{ab} = f^{acb} A^c
\ee
is the covariant derivative in the adjoint representation of the gauge group
and $\rho_\mathrm{ext}^a (\vx)$ denotes the ``external'' color charge density of the matter
fields. The operator $\Hat{D}\Pi$ on the left-hand side of Eq.~(\ref{G12}) is the generator
of time-independent small gauge transformations ($n[U]=0$). In the absence of external
charges, $\rho_\mathrm{ext}^a (\vx) = 0$, Gauss' law requires the wave functional to be invariant
under small gauge transformations only, while under large gauge transformation
it needs to be invariant only up to a phase \cite{JacReb76,CalDasGro76}
\be
\label{G14}
\Psi_\theta [A^U] = e^{- i \theta n [U]} \Psi_\theta [A] \; .
\ee
Here, $\theta$ is a free real parameter, which characterizes the vacuum wave
functional and is called the vacuum angle. Since $n [U] \in \mathbb{Z}$, the
wave functional of the $\theta$-vacuum Eq.~\eqref{G14} has the property 
\be
\label{G18}
\Psi_{\theta + 2 \pi} [A] = \Psi_\theta [A] \; ,
\ee 
which qualifies $\theta$ as an angle variable. $\theta$ is not known \textit{a priori} and
is not determined by the theory itself but must be fixed by experiment. One measurable effect
of $\theta$ would be a non-vanishing neutron electric dipole
moment \cite{Cre+79}. Current measurements \cite{Bak+06} restrict $\theta$ to the extremely small
value $| \theta | \leq 10^{- 10}$.

The transformation property Eq.~\eqref{G14} can be realized by the ansatz
\be
\label{G15}
\Psi_\theta [A] = e^{- i \theta W [A]} \phi[A] \; ,
\ee
where $\phi [A]$ is a gauge invariant wave functional, $\phi [A^U] = \phi [A]$,
and
\begin{align}
\label{G16}
W [A] &= \frac{g^2}{16\pi^2} \int \d[3]x \: K^0(x) \\
&= \frac{g^2}{16 \pi^2} \; \varepsilon_{ijk} \int \d[3]x \, \left[ A_i^a \partial_j A_k^a +
\frac{g}{3} \; f^{abc} A_i^a A_j^b A_k^c \right] \nonumber
\end{align}
is the Chern--Simons action, which changes under gauge transformations by the
winding number $n [U]$
\be
\label{G17}
W [A^U] = W [A] + n [U] \; .
\ee
The wave functional Eq.~\eqref{G15} does, however, not fulfill Eq.~\eqref{G18}. In the appendix we show
how Eq.~\eqref{G15} has to be modified to promote $\theta$ to an angle variable. We also show
there that Eq.~\eqref{G15} is a correct wave functional when $\theta$ is restricted to $[0,2\pi)$.
In the following, the cyclic property Eq.~\eqref{G18} of the vacuum wave functional will
be irrelevant, since we are anyway interested only in infinitesimally small $\theta$,
see Eq.~\eqref{chi} below. Therefore, we can use the simpler wave functional Eq.~\eqref{G15}.

The quantity of interest is the topological susceptibility,
Eq.~(\ref{G7}), which in the Hamiltonian approach can be defined by \cite{Wit79}
\be
\label{chi}
V \chi = \left. \frac{\d[2]\langle H \rangle_\theta}{\d\theta^2} \right|_{\theta=0}^{\textrm{no quarks}} \; ,
\ee
where $V$ is the spatial volume and
\be
\label{G27}
\langle H \rangle_\theta = \bra{\Psi_\theta} H  \ket{\Psi_\theta} \; , \quad \braket{\Psi_\theta}{\Psi_\theta} = 1
\ee
is the expectation value of the Yang--Mills Hamiltonian in the $\theta$-vacuum.
To evaluate the $\theta$-dependence of $\langle H \rangle_\theta$, the following identity will be useful
\be
\label{G28}
\frac{\delta W [A]}{\delta A^a_i (x)} = \frac{g^2}{8 \pi^2} \; B^a_i (x) \; ,
\ee
where $B_i^a$ is the magnetic field Eq.~(\ref{G11}). Obviously, the $\theta$-phase in
Eq.~(\ref{G15}) can
only contribute to the kinetic term of the Yang--Mills Hamiltonian Eq.~(\ref{G10}).
With Eq.~(\ref{G28}) we find from Eq.~(\ref{G15})
\be
\label{G29}
\Pi^a_i \Psi_\theta [A] = e^{- i \theta W [A]} \left( \Pi^a_i - \theta \,
\frac{g^2}{8 \pi^2} \; B^a_i \right) \phi [A] \; .
\ee
Inserting this relation into Eq. (\ref{G12}) and using the Bianchi identity
\be
\label{G30}
\hat{D}^{a b}_i B^b_i (x) = 0
\ee
we find that the wave functional $\phi [A]$ satisfies the same Gauss' law as
$\Psi_\theta [A]$
\be
\label{G31}
\hat{D}^{ab}_i \, \Pi^b_i \; \phi[A] = - g \, \rho_\mathrm{ext}^a \; \phi[A] \; .
\ee
To make contact with previous results obtained in the Hamiltonian approach to
Yang--Mills theory, see Refs.\ \cite{FeuRei04,EppReiSch07}, it is convenient to
work with the $\theta$-independent wave functional $\phi [A]$ and absorb the
$\theta$-dependence into the Hamiltonian by defining 
\be
\label{G32}
\bra{\Psi_\theta} H \ket{\Psi_\theta} = \bra{\phi} H_\theta \ket{\phi} \, .
\ee
Using Eq.~\eqref{G29} we find
\be
\label{G33}
H_\theta = \frac12 \int\left(\Pi_i^a  - \theta \, \frac{g^2}{8\pi^2} \; B_i^a \right)^2 + \frac12 \int \big( B_i^a \big)^2 .
\ee
An alternative way to arrive at this Hamiltonian is to add the topological
$\theta$-term directly to the original classical Lagrangian
\be\label{lag}
\begin{split}
\calL & = - \frac{1}{4} \; F^a_{\mu \nu} \, F^{\mu \nu}_a + \theta \, \frac{g^2}{32
\pi^2} \; F^a_{\mu \nu} \widetilde{F}^{\mu \nu}_a \\
& = \frac{1}{2} \big( \vec{E}^2 - \vec{B}^2 \big) + \theta \, \frac{g^2}{8 \pi^2} \; \vec{E} \cdot \vec{B} \; .
\end{split}
\ee
The $\theta$-term is a total derivative and does not contribute to the
classical equation of motion. It does, however, change the canonical momentum from $\Pi_i^a = F_{0i}^a$ to 
\be
\label{G37}
\Pi_i^a = F_{0i}^a + \theta \, \frac{g^2}{8 \pi^2} \; B_i^a
\ee
and after canonical quantization in Weyl gauge one finds again the Hamiltonian
$H_\theta$ (\ref{G33}).

Instead of working (for $\rho_\mathrm{ext}^a (\vx) = 0$) with gauge invariant wave functionals
$\phi [A]$, it is more convenient to explicitly resolve Gauss' law Eq.~(\ref{G31})
by fixing the gauge. For this purpose Coulomb gauge $\partial_i A_i^a = 0$
is particularly convenient and will be used in the following. In Coulomb gauge,
the gauge field is transversal $A = A^\perp$ but this is not true for the
momentum operator. We split the momentum operator into longitudinal and
transversal parts $\Pi = \Pi^\perp + \Pi^{||}$, where $\Pi^\perp = -i \delta/\delta A^\perp$.
The latter satisfies the canonical commutation relation for transversal fields
\be\label{comm:coulomb}
\comm{A_i^{\perp a}(\vx)}{\Pi_j^{\perp b}(\vy)} = i \, \delta^{ab} \, t_{ij}(\vx) \, \delta(\vx-\vy) \; ,
\ee
where $t_{ij}(\vx)=\delta_{ij} - \partial_i \partial_j/\partial^2$ is the transversal projector.

Gauss' law Eq.~(\ref{G31}) can be solved for the longitudinal part $\Pi^{||}$ in the standard fashion yielding
\be
\label{G38}
\Pi^{||} \ket{\phi} = - g \partial (- \hat{D} \partial)^{- 1} (\rho_\mathrm{ext} + \rho_g)
\ket{\phi} \; ,
\ee
where $\rho^a_g = \hat{A}^{\perp a b}_i \Pi^{\perp b}_i$ is the color charge density
of the gluons and $(- \hat{D} \partial)$ is the Faddeev--Popov kernel in Coulomb
gauge. With the aid of Eq.~(\ref{G38}), one derives from Eq.~(\ref{G33})
the gauge fixed Hamiltonian of the $\theta$-vacuum (by
considering $\langle \phi | H_\theta | \phi \rangle$ and using integration by parts
in the kinetic term). This yields\footnote{In the following, all field and momentum
operators are transversal and we will omit the symbol $\perp$.}
\be\label{ham}
H_\theta = H_0 + \theta \; \frac{g^2}{8\pi^2} \: H_1 + \left(\theta \; \frac{g^2}{8\pi^2} \right)^2 H_2 \; .
\ee
where
\begin{align}
H_0 &= \frac12 \int \d[3]x \: \left[ \calJ^{-1} \Pi_i^a(\vx) \calJ \Pi_i^a(\vx) + B_i^a(\vx) \, B_i^a(\vx) \right] + \nonumber \\
 & + \frac{g^2}{2} \int \d[3]x \: \d[3]y \: \calJ^{-1} \rho^a(\vx) \; \calJ \; F^{ab}[A](\vx,\vy) \rho^b(\vy)
\end{align}
is the usual Coulomb gauge fixed Hamiltonian \cite{ChrLee80} for $\theta=0$ and the $\theta$-dependent terms are given by
\begin{align}\label{ham:linear}
H_1 & = - \frac12 \int \! \d[3]x \left[ B^a_i(\vx) \Pi^a_i(\vx) +
\calJ^{-1} \Pi^a_i(\vx) \calJ B^a_i(\vx) \right] + \nonumber \\
& +  \frac12 \int \! \d[3]x \, \d[3]y \: \Big\{ G^{ab}[A](\vx,\vy) \partial_i^x B^a_i(\vx) g \rho^b(\vy) + \nonumber \\
& \, \quad + \calJ^{-1} g \rho^a(\vx) \calJ G^{ab}[A](\vx,\vy) \partial_i^y B^b_i(\vy) \Big\}
\end{align}
and
\be\label{ham:quadratic}
H_2 = \frac12 \int \d[3]x \: B_i^a(\vx) B_i^a(\vx) \; .
\ee
In the above expressions, \mbox{$\calJ=\Det(-\partial_i \Hat{D}_i)$} is the Faddeev--Popov determinant, $G$ is the Green's function of the Faddeev--Popov operator
\be
-\partial_i \Hat{D}_i^{ab}(\vx) \, G^{bc}[A](\vx,\vy)= \delta^{ac} \; \delta(\vx-\vy)\; ,
\ee
$F$ denotes the Coulomb operator
\be
F^{ab}[A](\vx,\vy) = \left[ (-\partial_i \Hat{D}_i)^{-1} (-\partial^2) (-\partial_j \Hat{D}_j)^{-1}\right]^{ab}_{\vx,\vy}
\ee
and $\rho$ is the sum of external and dynamical color charge densities
\be
\rho^a = \rho^a_\mathrm{ext} + \rho^a_g = \rho^a_\mathrm{ext} + f^{abc} A^b_i \, \Pi^c_i \; .
\ee
For the present purpose, the evaluation of the topological susceptibility $\chi$ (which
is entirely defined in the gluon sector), the external charges are not needed and we will put $\rho^a_\mathrm{ext}=0$
in the following. The Faddeev--Popov determinant $\calJ$ represents the Jacobian of the transformation
from the (flat) non gauge-fixed configuration space to the (curved) space of the Coulomb
gauge fixed fields. In particular, this Jacobian enters the integration measure of the
scalar product of wave functions
\be
(\Phi_1,\Phi_2) = \int \calD A \; \calJ \; \Phi_1^*[A] \; \Phi_2[A] \; .
\ee
The integration of transversal field configurations extends over the first Gribov region
$\Omega$ \cite{Gri78}, allowing the surface terms to be discarded in integration by parts.
Although the integration should be restricted to the fundamental modular region
$\Lambda \subset \Omega$ \cite{Zwa97}, there is evidence that integration over $\Omega$
yields the same expectation values \cite{Zwa04}. 

Following Ref.~\cite{FeuRei04} we introduce the ``radial'' wave functional
\be\label{radial:functional}
\widetilde{\Phi}[A] = \calJ^{1/2} \; \Phi[A]
\ee
which removes the Faddeev--Popov determinant from the integration measure of the scalar product. Matrix elements of observables $O[A,\Pi]$ can then be expressed as
\be\label{scal:prod}
\int \calD A \; \calJ \; \Phi^*_1 \, O \, \Phi_2 = \int \calD A \; \widetilde{\Phi}_1^* \, \widetilde{O} \, \widetilde{\Phi}_2 \; ,
\ee
where we have introduced the transformed operator
\be
\widetilde{O}[A,\Pi] = \calJ^{1/2} \; O[A,\Pi] \calJ^{-1/2} = O[A,\widetilde{\Pi}] \; .
\ee

An operator $O[A]$ depending only on the field variable $A_i^a(\vx)$ is obviously
not changed by this transformation, while the momentum operator transforms as
\be
\widetilde{\Pi}_i^a(\vx) = \Pi_i^a(\vx) + \frac{i}{2} \; \frac{\delta\ln\calJ}{\delta A_i^a(\vx)} \; .
\ee

The transformed Hamilton operator \mbox{$\widetilde{H}_\theta = \calJ^{1/2} H_\theta \calJ^{-1/2}$}
is obtained from $H_\theta$ by replacing $\Pi$ by $\widetilde{\Pi}$. $\widetilde{H}_0$
was explicitly given in Ref.~\cite{FeuRei04}. Furthermore, since $B_i^a(\vx)$ is a function of
the field variable only, $H_2$ does not change ($\widetilde{H}_2=H_2$). The explicit expression
for $\widetilde{H}_1$ is given in the next section.


\section{Matrix elements for the topological susceptibility}

We are interested here in the topological susceptibility, Eq.~\eqref{chi}. Since this quantity is defined as second derivative of $\langle H_\theta\rangle$ at $\theta=0$, it is sufficient to calculate $\langle H_\theta\rangle$ up to second order in $\theta$. This requires to treat $H_2$ in first order and $\widetilde{H}_1$ up to second order perturbation theory in $\theta$. We then find
\be\label{def:chi}
V \, \chi = 2 \left( \frac{g^2}{8\pi^2} \right)^2
\left[ \bra{0} H_2 \ket{0} - \sum_{n\neq0} \frac{|\bra{0}  \widetilde{H}_1 \ket{n}|^2}{E_n} \right]
\ee
where $\{| n \rangle\}$ denotes the set of eigenstates of $\widetilde{H}_0$. We use here a Dirac notation for the radial wave
functionals \eqref{radial:functional}, $\braket{A}{0}=\widetilde{\Phi}_0[A]$.

For the unperturbed Hamiltonian $\widetilde{H}_0$, we use the variational results obtained in Refs.~\cite{FeuRei04,EppReiSch07}. For the (radial) vacuum wave functional the following Gaussian form was chosen
\begin{align}\label{vacuum}
&\widetilde{\Phi}_0[A] = \braket{A}{0} \nonumber \\
&= \calN \exp \left[ -\frac12 \int \! \d[3]x \, \d[3]y \: A_i^a(\vx) \, \omega_{ij}(\vx,\vy) \, A_j^a(\vy) \right] ,
\end{align}
where $\omega_{ij}(\vx,\vy)=t_{ij}(\vx)\omega(\vx,\vy)$ and $\omega(\vx,\vy)$ is an integration kernel determined by minimization of the energy density. By means of Wick's theorem we can express expectation values of powers of field operators by the gluon propagator
\be\label{gluon:prop}
\bra{0} A_i^a(\vx) \, A_j^b(\vy) \ket{0} = \frac12 \, \delta^{ab} \, \omega_{ij}^{-1}(\vx,\vy) .
\ee
Note that $\omega(\vx,\vy)$ depends only on $|\vx-\vy|$ and by Eq.~\eqref{gluon:prop} its Fourier transform represents the single quasi-gluon energy.

The vacuum wave functional $\ket{0}$ Eq.~\eqref{vacuum} is annihilated $a_i^a(x)\ket{0}=0$
by the operator
\be\label{ann:op}
a_i^a = \frac{1}{\sqrt{2}} \left[ \omega_{ij}^{1/2} A_j^a + i \, \omega_{ij}^{-1/2} \Pi_j^a \right] ,
\ee
which is the annihilation operator of a quasi-gluon with energy $\omega$.
Here and in the following we suppress the explicit spatial dependence of the involved quantities and
include the spatial coordinates in the Lorentz indices, so that contracted
Lorentz indices imply integration over the spatial coordinate. The
corresponding creation operator reads
\be\label{cre:op}
a_i^{a\dagger} = \frac{1}{\sqrt{2}} \left[ \omega_{ij}^{1/2} A_j^a - i \, \omega_{ij}^{-1/2} \Pi_j^a \right] ,
\ee
and from Eq.~\eqref{comm:coulomb} follows that these operators satisfy the usual Bose commutation relation
\be
\comm{a_i^a(\vx)}{a_j^{b\dagger}(\vy)} = \delta^{ab} \, t_{ij}(\vx) \, \delta(\vx-\vy)
\ee
(temporarily restoring the explicit spatial dependence for clarity).

By repeated application of the creation operator $a^\dagger$ Eq.~\eqref{cre:op} on
the vacuum Eq.~\eqref{vacuum} a complete basis for the gluon Fock-space of quasi-gluons is generated
\be\label{basis}
\ket{n} = \calN_n \prod_{k=1}^n a_{i_k}^{a_k\dagger}(\vx_k) \ket{0} \; .
\ee
For a proper normalization of these quasi-particle states the additional
normalization factor $\calN_n$ is required when two or more indices take the same
value, which can occur due to the bosonic character of these quasi-particle excitations.

With a complete basis of the Yang--Mills Hilbert space at our dipsosal we can
explicitly carry out the perturbative calculations in Eq.~\eqref{def:chi}. The
expectation value of $H_2$ can be straightforwardly evaluated, yielding
\begin{align}\label{m1}
\bra{0} &H_2 \ket{0} = \frac{N_c^2-1}{2} \; V \int \dfr[3]{k} \: \frac{\vk^2}{\omega(\vk)} + \nonumber \\
+ & g^2 \; \frac{N_c(N_c^2-1)}{16} \: V \int \dfr[3]{k} \: \dfr[3]{q} \: \frac{3-(\uvk \cdot \uvq)^2}{\omega(\vk) \, \omega(\vq)}
\end{align}

The second term in Eq.~\eqref{def:chi} contains a sum over an infinite number of
states with an arbitrary large number of quasi-gluons and we have to resort to
some approximations. Motivated by the variational calculation for $\theta=0$
\cite{FeuRei04,EppReiSch07} we restrict ourselves to terms involving up to two loops
in the energy. Under this assumption, $\widetilde{H}_1$ (cf.\ Eq.~\eqref{ham:linear}) is given by
\begin{align}
&\widetilde{H}_1 = - \int \d[3]x \: \Pi_i^a(\vx) \, B_i^a(\vx) + \nonumber \\
&+ \int \d[3]x \: \d[3]y \: G^{ab}[A](\vx,\vy) \,
\partial_i^x B_i^a(\vx) \, g \Hat{A}^{bc}_j(\vy) \, \Pi^c_j(\vy) \label{ham:1} \; ,
\end{align}
and we can use the following factorization in the matrix elements of the second term of Eq.~\eqref{ham:1}
\begin{align}\label{factorization}
\bra{0} G[A] \, &(\partial B) \,(g \Hat{A} \, \Pi) \ket{n} \nonumber \\ 
&\simeq \bra{0} G[A] \ket{0} \bra{0} (\partial B) \,(g \Hat{A} \, \Pi) \ket{n} \; ,
\end{align}
where $\langle G[A] \rangle$ is the ghost propagator. With these approximations, the relevant
contributions to the second term in Eq.~\eqref{def:chi}
\be\label{m2+m3}
- \sum_{n\neq0} \frac{|\bra{0}  \widetilde{H}_1 \ket{n}|^2}{E_n} =: \calM_2 + \calM_3
\ee
come from two quasi-gluon states, $n=2$
\be\label{m2:def}
\calM_2 = -\frac{1}{2!} \sum_{1,2} \frac{|\bra{0} B \Pi \ket{2} - \langle G \rangle \bra{0} \partial B g\Hat{A}\Pi \ket{2}|^2}{E_2} \; ,
\ee
and from the 3-quasi-gluon states, $n=3$
\be\label{m3:def}
\calM_3 = -\frac{1}{3!} \sum_{1,2,3} \frac{|\bra{0} B \Pi \ket{3}|^2}{E_3} \; .
\ee
The prefactors $1/2!$ and $1/3!$ take the normalization of the states \eqref{basis}
into account and avoid multiple counting. $E_2$ and $E_3$, respectively, denote the
energies of the two- and three-quasi-gluon states and accordingly are given by sums
of, respectively, two and three $\omega$'s.

The matrix elements in Eqs.~\eqref{m2:def}, \eqref{m3:def} can be evaluated by means of Wick's theorem, yielding
\begin{widetext}
\begin{align}
\calM_2 & = - \frac{N_c^2-1}{2} \: V \int \dfr[3]{k} \: \frac{\vk^2}{\omega(\vk)} + \nonumber \\
& - g^2 \; \frac{N_c(N_c^2-1)}{2} \: V \int \dfr[3]{k} \dfr[3]{q} \: \frac{\vk^2}{\omega(\vk)} \; G(\vk+\vq)
\left( \frac{1}{\omega(\vk)} + \frac{1}{\omega(\vq)} \right) \left[ \frac{1+(\uvk \cdot \uvq)^2}{2} + \frac{\vk\cdot\vq}{\vk^2} \right]
\label{m2} \\
\calM_3 &= - g^2 \; \frac{N_c(N_c^2-1)}{48} \: V \int \dfr[3]{k} \dfr[3]{q} \dfr[3]{p} \: 
\frac{\omega(\vk)+\omega(\vq)+\omega(\vp)}{\omega(\vk) \, \omega(\vq) \, \omega(\vp)}
\left[ 2 - 2 (\uvk\cdot\uvq)(\uvk\cdot\uvp)(\uvp\cdot\uvq) \right] (2\pi)^3 \delta(\vec{k}+\vec{q}+\vec{p}) \; \label{m3}.
\end{align}
\end{widetext}

Within our approximations the expression in the bracket in Eq.~\eqref{def:chi}
is given by the sum of Eqs.~\eqref{m1}, \eqref{m2} and \eqref{m3}.
The term involving $\langle G \rangle ^2$ in Eq.~\eqref{m2:def} does not appear in
Eq.~\eqref{m2} since it involves three loops. It is worth
noting that the first term in Eq.~\eqref{m2} cancels the first term in Eq.~\eqref{m1}.
This cancellation is due to the fact that the Gaussian wave functional is exact in
an Abelian theory and the two terms are actually the leading order contribution to
the topological susceptibility in perturbation theory in powers of $g$,
where $\chi$ vanishes identically. In Ref.~\cite{Cam06} it has
been explicitly shown that this cancellation occurs also in next-to-leading order.

The integrals in Eqs.~\eqref{m1}, \eqref{m2} and \eqref{m3} are UV divergent.
Since the topological susceptibility $\chi$ vanishes to any (finite) order
perturbation theory in $g$, in principle, all UV-divergences should cancel. However,
the approach \cite{FeuRei04gb,FeuRei04} we are using is nonperturbative and, due
to the approximations involved, does not include all terms of a given power
in $g$. As a consequence, there is a mismatch of UV-singularities and our
expression for $\chi$ is UV-divergent. To remove these spurious UV-singularities
we subtract from all propagators the corresponding tree-level form,
\textit{i.e.}, we make the following replacements
\begin{subequations}\label{subtraction}
\begin{align}
\omega^{-1}(\vk)& \rightarrow \omega^{-1}_s(\vk) = \omega^{-1}(\vk)-1/\sqrt{\vk^2} \; , \\
G(\vk) & \rightarrow G_s(\vk) = G(\vk)-1/\vk^2 \; .
\end{align}
\end{subequations}
This is in the spirit of the zero-momentum subtraction scheme.
The replacement \eqref{subtraction} makes the integrals \eqref{m1}, \eqref{m2} and \eqref{m3} convergent.
Using the symmetry of the integrands we can finally cast the expression for the topological
susceptibility into the form
\be\label{chi:result}
\chi= g^2 \left( \frac{g^2}{8\pi^2} \right)^2 N_c(N_c^2-1) ( I_1 + I_2 ) \, ,
\ee
where
\begin{widetext}
\begin{subequations}\label{integrals}
\begin{align}
I_1 &= \frac{1}{4} \int \dfr[3]{k} \: \dfr[3]{q} \: \omega^{-1}_s(\vk) \: \omega^{-1}_s(\vq) \:
\left[ 1 - \vk^2 \: \frac{1-(\uvk \cdot \uvq)^2}{(\vk+\vq)^2} \right] \label{I1} \\
I_2 &= - \int \dfr[3]{k} \: \dfr[3]{q} \: k^2 \: \omega^{-1}_s(\vk) \: \left[ \omega^{-1}_s(\vk)+\omega^{-1}_s(\vq) \right]
G_s(\vk+\vq) \left[ \frac{1+(\uvk \cdot \uvq)^2}{2}+\frac{\vq \cdot \vk}{\vk^2} \right] . \label{I2}
\end{align}
\end{subequations}
\end{widetext}

Eq.~\eqref{I1} follows from the sum of Eq.~\eqref{m3} and the second term in Eq.~\eqref{m1}
while Eq.~\eqref{I2} follows from the second term in Eq.~\eqref{m2}. The angular integration
in Eq.~\eqref{I1} can be performed analytically.


\section{Results}

The expressions for the topological susceptibility, Eqs.~\eqref{chi:result}
and \eqref{integrals}, depend via the gluon and ghost propagators,
$\omega^{-1}$ and $G$, on the vacuum properties, \textit{i.e.}, on the vacuum wave functional.
We will use here the ghost and gluon propagators determined in Ref.~\cite{EppReiSch07} as input.
Furthermore, to simplify our calculations the numerical solutions obtained in Ref.~\cite{EppReiSch07}
for the gluon and ghost propagators were fitted by the following ans\"atze
\begin{subequations}\label{propagators}
\begin{align}
\omega(\vk) &= \sqrt{\vk^2 + \frac{m^4}{\vk^2} } \; ,  \label{gluon:gribov} \\
G(\vk) &= \frac{1}{\vk^2} \sqrt{ 1 + \frac{M^2}{g^2 \, \vk^2} } \; , \label{ghost:gribov}
\end{align}
\end{subequations}
($g$ is the coupling constant).
\begin{figure}
\includegraphics[width=.9\columnwidth]{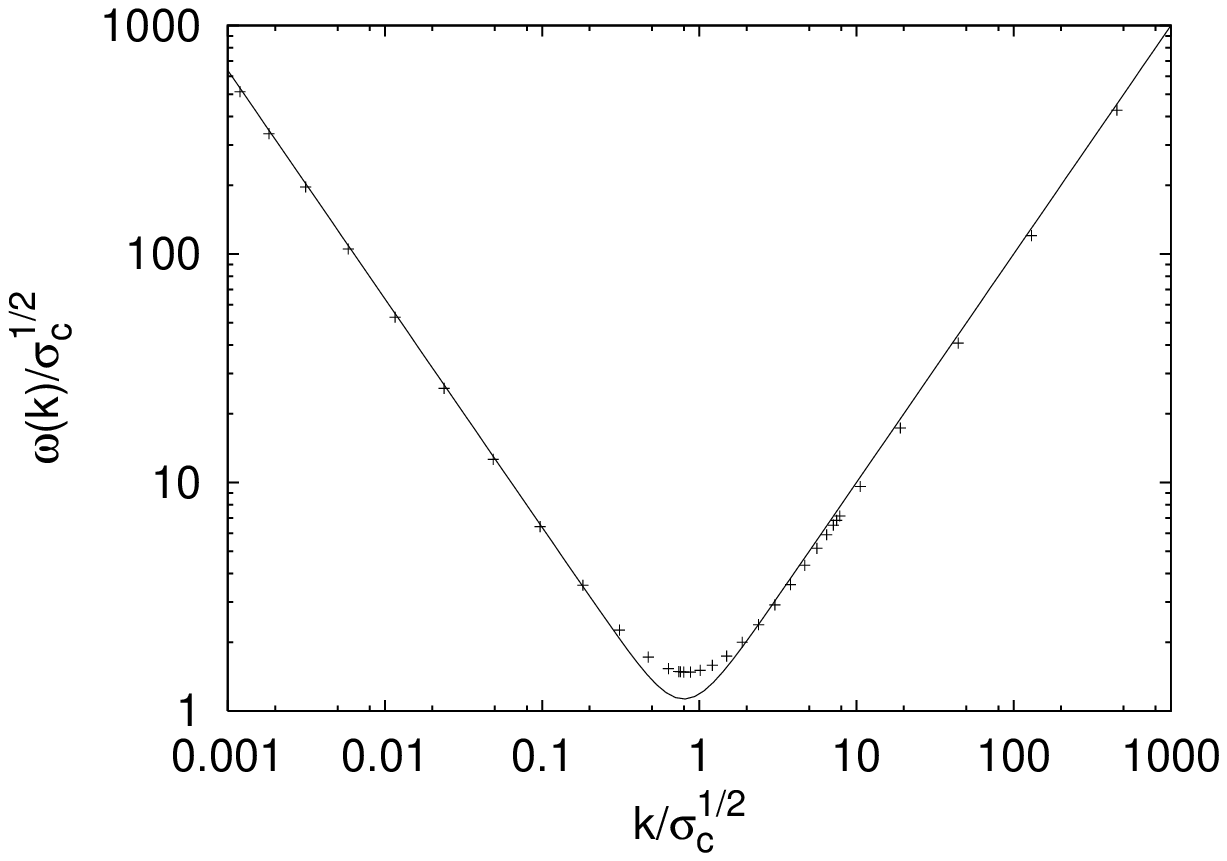}
\includegraphics[width=.9\columnwidth]{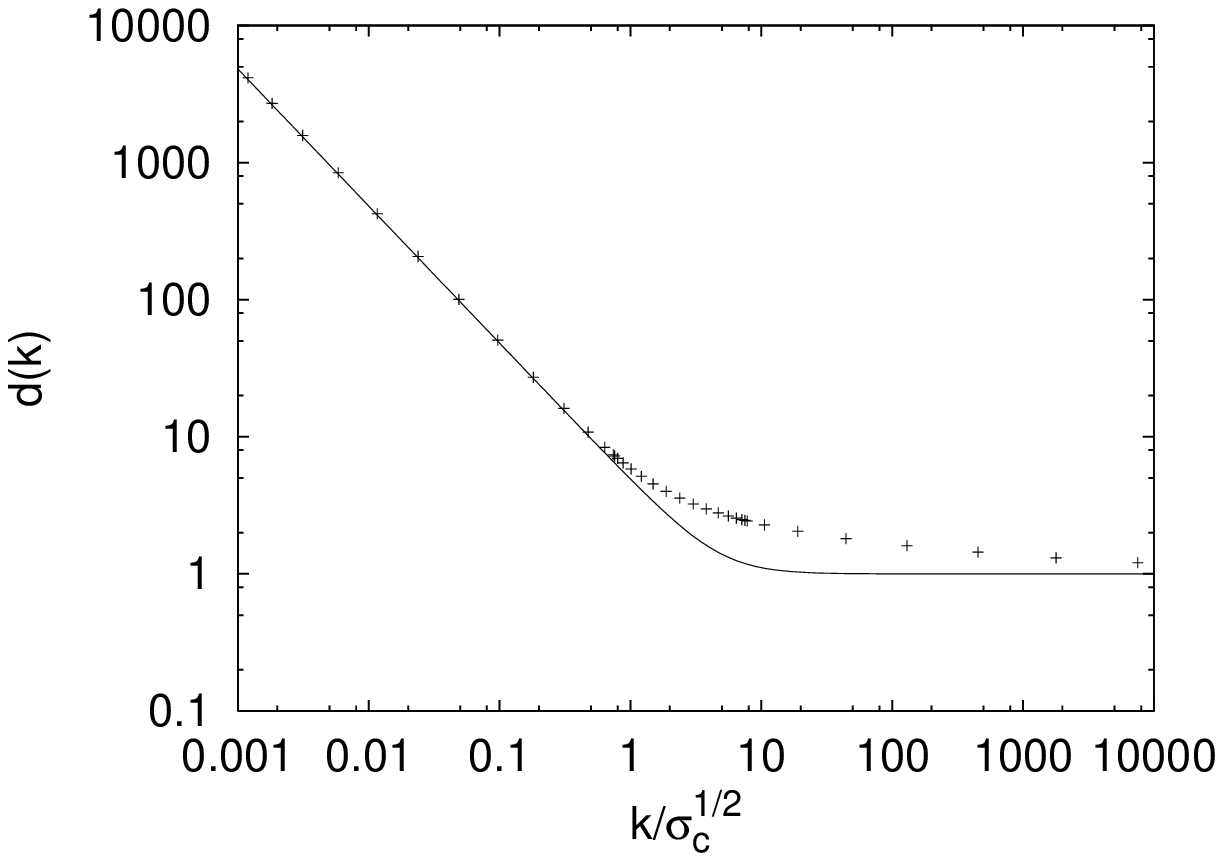}
\caption{\label{fig:propagators} Comparison between the numerical results of Ref.~\cite{EppReiSch07} (crosses) and the parametrizations \eqref{propagators} (lines) for the inverse gluon propagator (top panel) and the ghost form factor (bottom panel).}
\end{figure}
The parametrization \eqref{gluon:gribov} of the gluon energy was already
assumed heuristically by Gribov \cite{Gri78}. A factor $g$ was included in
the parametrization of the ghost propagator \eqref{ghost:gribov} in order
to facilitate the fitting to the numerical results of Ref.~\cite{EppReiSch07},
where a factor $g$ was included in the definition of the ghost form factor $d(\vk)$
\footnote{The inverse ghost form factor $g d^{-1}(\vk)$ has the meaning of the dielectric function of the Yang--Mills vacuum \cite{Rei08}.}
\be\label{ghost:form:factor}
G(\vk)=\frac{d(\vk)}{g \, \vk^2} \; .
\ee
The numerical solutions of Ref.~\cite{EppReiSch07} and their pa\-ra\-me\-trizations by
Eq.~\eqref{propagators} are shown in Fig.~\ref{fig:propagators}. The integrals Eq.~\eqref{integrals}
entering the topological susceptibility Eq.~\eqref{chi:result} recieve their dominant
contributions from the infrared momentum regime, where the parametrizations Eq.~\eqref{propagators}
give a perfect fit to the numerical solutions. The infrared part
of the parametrizations Eq.~\eqref{propagators}
\be\label{IR:propagators}
\omega_\mathrm{IR}(\vk) = \frac{m^2}{\sqrt{\vk^2}} \; , \quad d_\mathrm{IR}(\vk) = \frac{M}{\sqrt{\vk^2}}
\ee
was fitted to the infrared behaviour of the gluon and ghost propagators found in
Ref.~\cite{EppReiSch07} by solving the corresponding Dyson--Schwinger equations (DSEs).
In the Hamiltonian approach in Coulomb gauge the physical scale is the Coulomb
string tension $\sigma_c$, \textit{i.e.}, the coefficient of the linear term in the non-Abelian Coulomb
potential. In units of the Coulomb string tension $\sigma_c$ the fit of Eq.~\eqref{IR:propagators} to the numerical
solutions of the DSEs yields $m^2=0.614$ and $M=4.97$, while from the (analytic)
infrared analysis of the DSEs \cite{FeuRei04,SchLedRei06,EppReiSch07} one extracts the values $m^2 = 2/\pi \simeq 0.637$
and $M = \sqrt{8\pi} \simeq 5.01$.

\begin{figure}
\includegraphics[width=.9\columnwidth]{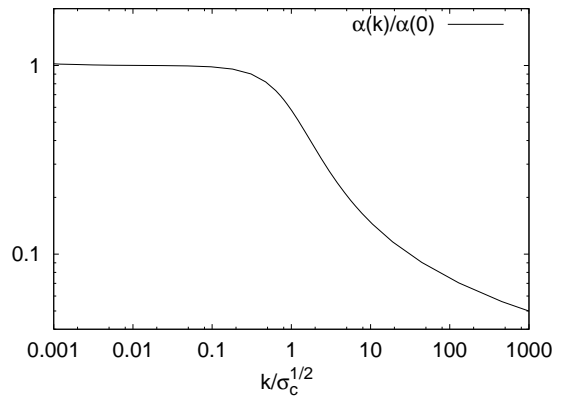}
\caption{\label{coupling}Running coupling as calculated in \cite{EppReiSch07}.}
\end{figure}

With the definition Eq.~\eqref{ghost:form:factor} of the ghost form factor in the variational
calculation of Refs.~\cite{FeuRei04,FeuRei04gb,ReiFeu05,EppReiSch07,SzcSwa01,Szc04} the coupling constant $g$ drops out
from the DSEs and the value of the coupling constant never had to be
specified. Contrary to this, the topological susceptibility Eq.~\eqref{chi:result} explicitly
contains the coupling constant. In principle, after a complete renormalization procedure, in all physical
(renormalized) quantities the coupling constant should be replaced by the running one defined
in a renormalization group invariant way. Such a complete
renormalization program is not yet feasible, although some progress in this direction has been made\footnote{The counter terms required for the renormalization have been identified \cite{Epp+07,ReiCam08}.}. Fortunately the running couling constant calculated
in Ref.~\cite{EppReiSch07} in the Hamiltonian approach in Coulomb gauge (see Fig.~\ref{coupling}) has a very weak momentum
dependence in the infrared regime. It basically stays constant below the infrared scale $\sqrt{\sigma_c}$.
It is this low-momentum regime which gives the major contribution to the integrals Eq.~\eqref{integrals}.
We will therefore use the plateau value of the running coupling constant.
Using the definition of the nonperturbative running coupling given in Ref.\ \cite{FisZwa05},
its value at zero momentum is \cite{SchLedRei06}
\be
\alpha_s(0) = \frac{g^2(0)}{4\pi} = \frac{16 \pi}{3 N_c} \; .
\ee
Numerical (Gauss-Legendre) evaluation of the integrals Eq.~\eqref{integrals} yields
\be
I_1 = (0.077 \sqrt{\sigma_c})^4 , \quad I_2 = (0.021 \sqrt{\sigma_c})^4.
\ee
Then we find for the topological susceptibility for $N_c=2$
\be\label{result}
\frac{\chi^\frac14}{\sqrt{\sigma_c}} = 0.45 \; .
\ee

\begin{figure}
\includegraphics[width=.9\columnwidth]{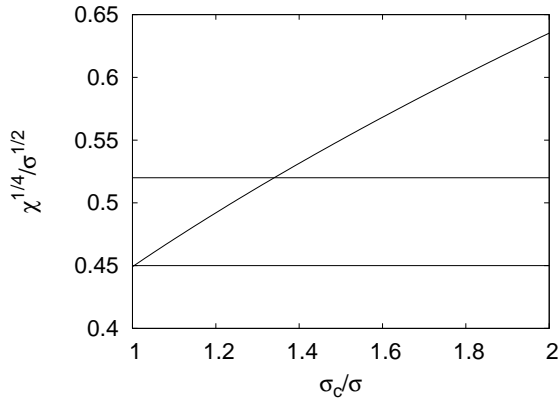}
\caption{\label{chi.bild}Value of $\chi^\frac14/\sqrt\sigma$ depending on the ratio $\sigma_c / \sigma$. The two horizontal lines limit the range of the lattice results.}
\end{figure}

It was shown in Ref.~\cite{Zwa03b} that $\sigma_c$ is an upper bound for the string tension $\sigma$
extracted from the Wilson loop. Lattice calculations performed in Ref.~\cite{LanMoy04} for $SU(2)$ show
that $\sigma_c \simeq 1.5 \, \sigma$, while Ref.~\cite{GreOleZwa04} seems to suggest an even larger value
for $\sigma_c$. In Fig.~\ref{chi.bild} we present our numerical result for the topological
susceptibility as a function of $\sigma_c/\sigma$. For $1< \sigma_c/\sigma<1.33$, $\chi$ is inside the range
predicted by the lattice data \cite{Tep97,Tep99,deFor+97,DeG+97}.
Choosing $\sigma_c=1.5\,\sigma$ we find with \mbox{$\sqrt{\sigma}=440$ MeV}
\be
\chi = ( 240 \mbox{ MeV} )^4 .
\ee
This value is somewhat larger than the lattice prediction $\chi=(200-230$ MeV$)^4$.


\section{Summary and conclusions}

We have calculated the topological susceptibility $\chi$ within the Hamiltonian approach to
Yang--Mills theory in Coulomb gauge using the gluon and ghost propagators determined
previously by a variational solution of the Yang--Mills Schr\"odinger equation as input. To our knowldedge
this is the first \emph{ab initio} continuum calculation of the topological susceptibility,
granted the approximations adopted. The precise numerical value for $\chi$ depends on the relation
between the Coulomb string tension and the one extracted from the Wilson loop. It is therefore
very desirable to perform a sophisticated calculation of the Wilson loop within the present
approach. Adopting the relation $\sigma_c = 1.5 \, \sigma$ as suggested by lattice calculations,
the numerical value obtained for $\chi$ is in reasonable agreement with the lattice data. The
results obtained in the present paper are quite encouraging for the calculation of further
hadronic quantities within the present approach. This will require the inclusion of the quarks. 


\begin{acknowledgments}
The authors would like to thank Jan Pawlowski, Lorenz von Smekal, Axel Weber and Daniel Zwanziger for valuable discussions.
They are also indebted to Giuseppe Burgio, Markus Quandt, Wolfgang Schleifenbaum and Peter Watson for a
critical reading of the manuscript and useful comments. We also would like to thank Adam Szczepaniak for
continuing discussions on the Hamiltonian approach to QCD in Coulomb gauge.
This work was supported by the Deutsche Forschungsgemeinschaft (DFG) under contracts No.~Re856/6-1
and No.~Re856/6-2 and by the Cusanuswerk--Bisch\"ofliche Studienf\"orderung. 
\end{acknowledgments}


\appendix*
\section{}

In the following we show how the wave functional Eq.~\eqref{G15} can be modified to fulfill
Eq.~\eqref{G18}, so that $\theta$ becomes a true angle variable. For this purpose consider
the wave functional
\be
\label{G19}
\bar{\Psi}_\theta [A] = \int^{2 \pi}_0 \d\theta' \: \delta [\theta, \theta'] \: \Psi_{\theta'} [A] \; ,
\ee
where 
\be
\label{G20}
\delta [\theta, \theta'] = \frac{1}{2 \pi} \sum_m e^{- i m (\theta - \theta')}
\ee
is the periodic $\delta$-function satisfying $\delta [\theta + 2 \pi , \theta']
= \delta [\theta, \theta']$. Inserting Eq.~(\ref{G15}) into Eq.~(\ref{G19}) we obtain
\be
\label{G21}
\bar{\Psi}_\theta [A] = \sum_m e^{- i  \theta m} f (m - W [A]) \; \phi [A] \; ,
\ee
where
\be
\label{G22}
f (x) = \frac{1}{2 \pi} \int^{2 \pi}_0 \d\theta' \: e^{i \theta' x} \; .
\ee
The functional $\bar{\Psi}_\theta [A]$  satisfies both Eq.~(\ref{G14}) and Eq.~(\ref{G18}) and
thus can be used as the wave functional of the $\theta$-vacuum. According to Eq.~\eqref{G21}, it has the form
\be
\label{G23}
\bar{\Psi}_\theta [A] = \sum_m e^{- i \theta m} \, \Psi_m [A] \; ,
\ee
where
\be
\label{G24}
\begin{split}
\Psi_m [A] &= f (m - W [A]) \; \phi [A] \\
&= \frac{1}{2\pi} \int_0^{2\pi} \d\theta' \; e^{im\theta'} \; \Psi_{\theta'}[A]
\end{split}
\ee
is a localized wave functional centered at the classical vacuum $U \partial
U^\dagger$ with winding number $n [U] = m$. It satisfies the relation
\be
\label{G25}
\Psi_m [A^U] = \Psi_{m - n [U]} [A] \; .
\ee
Replacing $\bar{\Psi}_\theta [A]$ by $\Psi_m [A]$ corresponds to the so-called
``tight binding'' approximation in solid states physics.

We are not using this approximation. From Eq.~\eqref{G24} it is seen that our wave
functional $\Psi_\theta[A]$ is the Fourier transform of $\Psi_m[A]$.

Using the Poisson relation, the periodic $\delta$-function Eq.~\eqref{G20} can
be expressed as
\be
\delta[\theta,\theta'] = \sum_{l=-\infty}^{\infty} \delta(\theta - \theta' + 2 \pi l) \; ,
\ee
where $\delta(x)$ denotes the ordinary $\delta$-function. Inserting this representation
into Eq.~\eqref{G19} we find
\be
\bar{\Psi}_\theta [A] = \Psi_{\theta+2\pi l_0}[A] \; ,
\ee
where $l_0$ is defined by the condition
\be
\theta + 2 \pi l_0 \in [0,2\pi) \, .
\ee
This shows that $\Psi_\theta[A]$, Eq.~\eqref{G15}, is a correct representation of the
wave functional of the $\theta$-vacuum for $\theta \in [0,2\pi)$.


\bibliography{biblio}{}
\bibliographystyle{h-physrev3}

\end{document}